\documentclass[notitlepage,10pt]{article}

\input{psfig.sty}
\begin{document}

\title{Geometric Parameterization of $J/\Psi$ Absorption in Heavy Ion Collisions}

\author{
M. J. Bennett\\Physics Division, Los Alamos National Laboratory\\Los Alamos, NM 87545 
\\[0.7cm]
J. L. Nagle\\
Columbia University, Nevis Laboratory\\Irvington, NY 10533}
\date{November 24, 1998}
\maketitle

\begin{abstract}
We calculate the survival probability of $J/\Psi$ particles in various 
colliding systems using a Glauber model.  An analysis of recent data\cite{na50} 
has reported a $J/\Psi$-nucleon breakup cross section of 6.2$\pm$0.7 mb\cite{na50_qm97} 
derived from an exponential fit to the ratio of $J/\Psi$ to Drell-Yan yields as a function 
of a simple, linearly-averaged mean path length $\langle L \rangle$ 
through the nuclear medium.  Our calculations indicate that, 
due to the nature of the calculation, this approach yields an apparent 
breakup cross section which is systematically lower than the actual value.  

\end{abstract}

\section{Introduction}

Recently, data from the NA50 collaboration on $J/\Psi$ yields from p+A and A+A 
collisions have become available, which have been interpreted as displaying 
``anomalous $J/\Psi$ suppression''\cite{na50}.  It has been predicted that in the
case of a phase transition to a quark-gluon plasma (QGP), the yield of $J/\Psi$ 
is suppressed due to Debye screening\cite{mat_satz}.  There have been numerous efforts 
to explain this data, invoking such effects as interactions with co-moving 
hadrons\cite{gavin} and suppressed production due to initial state energy 
loss\cite{frankel,mjb_jln_energy}, as well as decreased yields in a QGP\cite{wong}.  

However, even in the absence of a deconfinement transition, the $J/\Psi$ can 
undergo inelastic interactions as it traverses the nuclear medium (as has been 
shown recently in data from p+A collisions\cite{pa}), which reduces the overall 
$J/\Psi$ yield.  Thus, the signature for a deconfinement transition would be 
$J/\Psi$ suppression at a level beyond that expected from the normal nuclear 
medium, an effect referred to as ``anomalous $J/\Psi$ suppression''\cite{na50}.

In the analysis of the NA50 data, the ratio of observed $J/\Psi$ to Drell-Yan 
yields is interpreted as proportional to the $J/\Psi$ 
survival probability.  This ratio is then fit as an exponential function of a simple, 
linearly-averaged mean path length through the nuclear medium $\langle L \rangle$
which returns a value for 
the $J/\Psi$-nucleon breakup cross section of 6.2$\pm$0.7 mb\cite{na50_qm97}.  The
comparison of this experimentally determined cross section and theoretical modeling is
important in determining the relative contributions 
of the ``pre-resonance'' color octet and the color singlet 
states\cite{sig_calc,kharz_zphys,qiao,boris}
to the overall absorption in A-A collisions, as well as in estimating the magnitude 
of the anomalous $J/\Psi$ suppression. 

In this paper we describe a calculation of $J/\Psi$ survival probability in A+A collisions.  
The results of this calculation indicate that a breakup cross section derived in the manner 
of \cite{na50} is systematically lower than the actual value.  We will discuss the origin 
of this discrepancy.
                      
\section{The Geometric Model}

In order to study the effect of collision geometry on the $J/\Psi$ survival probability, 
we have utilized a Glauber model approach, implementing Woods-Saxon parameterizations 
of nucleon distributions within the colliding nuclei.
Thus, assuming spherical symmetry for the nuclei, the nucleon density is described by:
\begin{equation}
\rm{{{dN} \over {dR}} \propto \left[ 1 + exp \left( {{R - C} \over {D}}\right) \right]^{-1}}
\end{equation}
where 
\begin{equation}
\rm{C = R_{0} \times A_{nucleus}^{1/3}}
\end{equation}
where $\rm{R_{0}}$= 1.11 fm, $\rm{A_{nucleus}}$ is the atomic number of the nucleus, 
and D = 0.75 fm.  For the specific case of a Pb nucleus, the parameters used are 
$\rm{R_{0}}$ = 1.1 fm, C = 6.624 fm, and D = 0.549 fm, as given in \cite{woods-saxon}.
The nucleon-nucleon cross section used in the Glauber calculation is $\sigma_{int}$=30 mb.
The results of the calculation are not sensitive to the exact value of this parameter 
over the range 25-40 mb.

Production of $J/\Psi$ is modeled by assuming that each nucleon-nucleon 
collision has the same probability of producing a $J/\Psi$.  We note that 
this modeling assumes that initial state energy 
loss does not affect $J/\Psi$ production.  We are currently investigating the validity of this condition; the results of this study will 
be addressed in detail in a follow-up paper\cite{mjb_jln_energy}.

After production, the $J/\Psi$ is taken to be at rest in the center-of-mass frame, 
such that the survival probability is a function of the path length through 
nuclear material which the $J/\Psi$ must traverse, the nucleon density and 
the breakup cross section $\sigma_{J/\Psi-N}$.
Specifically, the survival probability calculation requires an integration over $z$:
\begin{equation}
P_{surv}^{tgt,proj}=\int_{z_{production}}^{z_{tgt,proj}}\exp(-\rho(R)\: \sigma_{J/\Psi-N}z)\:dz
\end{equation}
where $z_{tgt,proj}$ is the position of the trailing edge of the target and 
projectile nuclei, respectively (defined by the $z$ value, at the same $x$ 
and $y$ values as the production point, where the nucleon density is negligibly 
small) and $\rho(R)$ is the local nucleon density.  The total survival 
probability is just the product of the two terms, $P_{surv}=P_{surv}^{tgt} \times P_{surv}^{proj}$.

Clearly, the survival probability of a given $J/\Psi$ depends strongly on where 
in the interaction region it is produced.  However, for any given collision 
geometry ({\bf i.e.} impact parameter), there are a wide range of possible production points, each with a 
different $J/\Psi$ survival probability. Thus, the mean survival probability 
for a given geometry represents an 
average over all possible $J/\Psi$ production points within the interaction 
region.  We note that in the interpretation of actual experimental data, the 
situation is further complicated by the impossibility of experimentally selecting 
on impact parameter, given finite detector resolution and the natural dispersion 
in the various centrality-related measurable quantities.  Thus, measured $J/\Psi$ 
yields represent an average over various collision geometries in addition to the 
average over survival probabilities for a fixed, ``ideal'' geometry, as discussed here.

In previous analyses\cite{na50}, the $J/\Psi$ survival probability was taken as 
proportional to the ratio of observed $J/\Psi$ yields to Drell-Yan.  For various 
classes of event geometries, this ratio was fit as an exponential function of 
$\langle L \rangle$, the mean path length through the nuclear medium which the 
$J/\Psi$ must traverse.  In that approach, $\langle L \rangle$ for a given class 
of geometries was the average of $\overline{L}(b)$ for the impact parameters $b$ 
contained in that class and
\begin{equation}
\overline{L}(b)=\frac{1}{\rho_{0}}\left\langle \int \rho \, dz \right\rangle,
\end{equation}
where $\rho_{0}$ is the mean nucleon density, $\rho$ is the local nucleon density 
and the integration is performed over the interaction region.  We have performed 
a similar calculation within our Glauber model.  In our model, we have used a
mean nuclear density of $\rho_{0}$=0.1793 nucleons/fm$^3$; the results of the calculation are insensitive to this normalization factor.  For a given colliding system, the 
value $\langle L \rangle$ averaged over all impact parameters ({\em i.e.} minimum 
bias) can be calculated analytically (see \cite{na50} for details), which provides 
a cross check of the $\langle L \rangle$ calculation; our values agree well with 
the expected values.

\begin{center}
\begin{figure}[tb]
\psfig{file=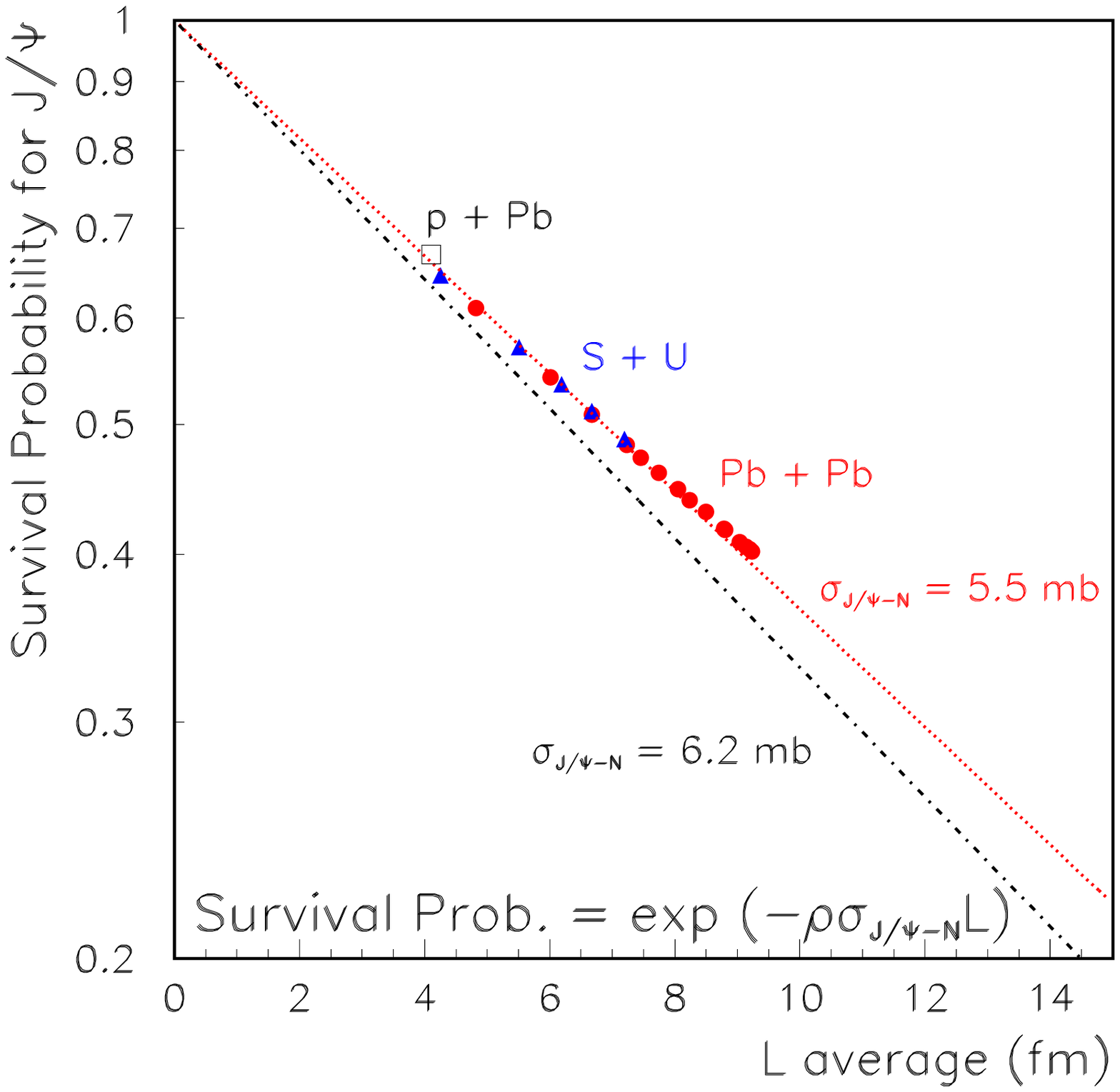,width=4.0in,bbllx=0pt,bblly=140pt,bburx=560pt,bbury=680pt,clip=}
\caption{Simulated $J/\Psi$ survival probability as a function of $\langle L \rangle$, the mean path length through the nuclear medium, for various colliding systems (p+Pb are open squares, S+U are solid triangles and Pb+Pb are solid circles--see the text for details on the values of $\langle L \rangle$).  The points were generated using a breakup cross section of 6.2 mb in the nuclear medium.  A fit to the calculated survival probabilities, shown as a dotted line, returns an effective absorption cross section of 5.5 mb.  For comparison, the dashed line indicates the trend the points should take if $\langle L \rangle$ were a valid indicator of the mean free path in the nuclear medium.}
\label{fig:lave}
\end{figure}
\end{center}

Shown in Fig.~\ref{fig:lave} are the calculated $J/\Psi$ survival probabilities for 
various colliding systems and geometries, plotted as a function of $\langle L \rangle$.  
To arrive at the specific $\langle L \rangle$ points given in the figure, we have 
simulated NA50 transverse energy bins\cite{na50_qm97}; transverse energy is assumed 
to scale as the number of wounded nucleons\cite{kharz_qm96}.  However, the substance 
of the figure does not depend on the specific centrality bins chosen.  For the points 
shown, a breakup cross section of  
$\sigma_{J/\Psi-N}$ = 6.2 mb is used as input to our Glauber calculation.  As can be 
seen clearly in Fig.~\ref{fig:lave}, the apparent breakup cross section of 5.5 mb 
obtained from an exponential fit versus $\langle L \rangle$ differs substantially 
from the actual breakup cross section.  
Further, the points from the calculation, while they give a reasonable fit, do not lie 
precisely on an exponential.
Since the points do not lie on a perfect exponential, the amount of error incurred by fitting 
as a function of $\langle L \rangle$ will depend on the exact centrality binning 
implemented.  However, we emphasize that the discrepancy between the apparent and 
actual breakup cross sections is inherent in the collision geometry, and would be 
present even if it were possible experimentally to precisely determine the event 
impact parameter.

It is important to gauge the level of the deviation from exponential simply
due to the error in calculating $\langle L \rangle$, since the
anomalous suppression in Pb+Pb collisions is measured relative to an extrapolation of the exponential fit.  
We have found that there is a systematic deviation from exponential, 
but of a relatively small magnitude.  And it is notable that this deviation for 
the most central Pb + Pb values is above the projected exponential fit, 
in the opposite direction of the suppression.  Thus the observed suppression is
not explained away by the error in plotting versus $\langle L \rangle$.

\begin{center}
\begin{figure}[tb]
\psfig{file=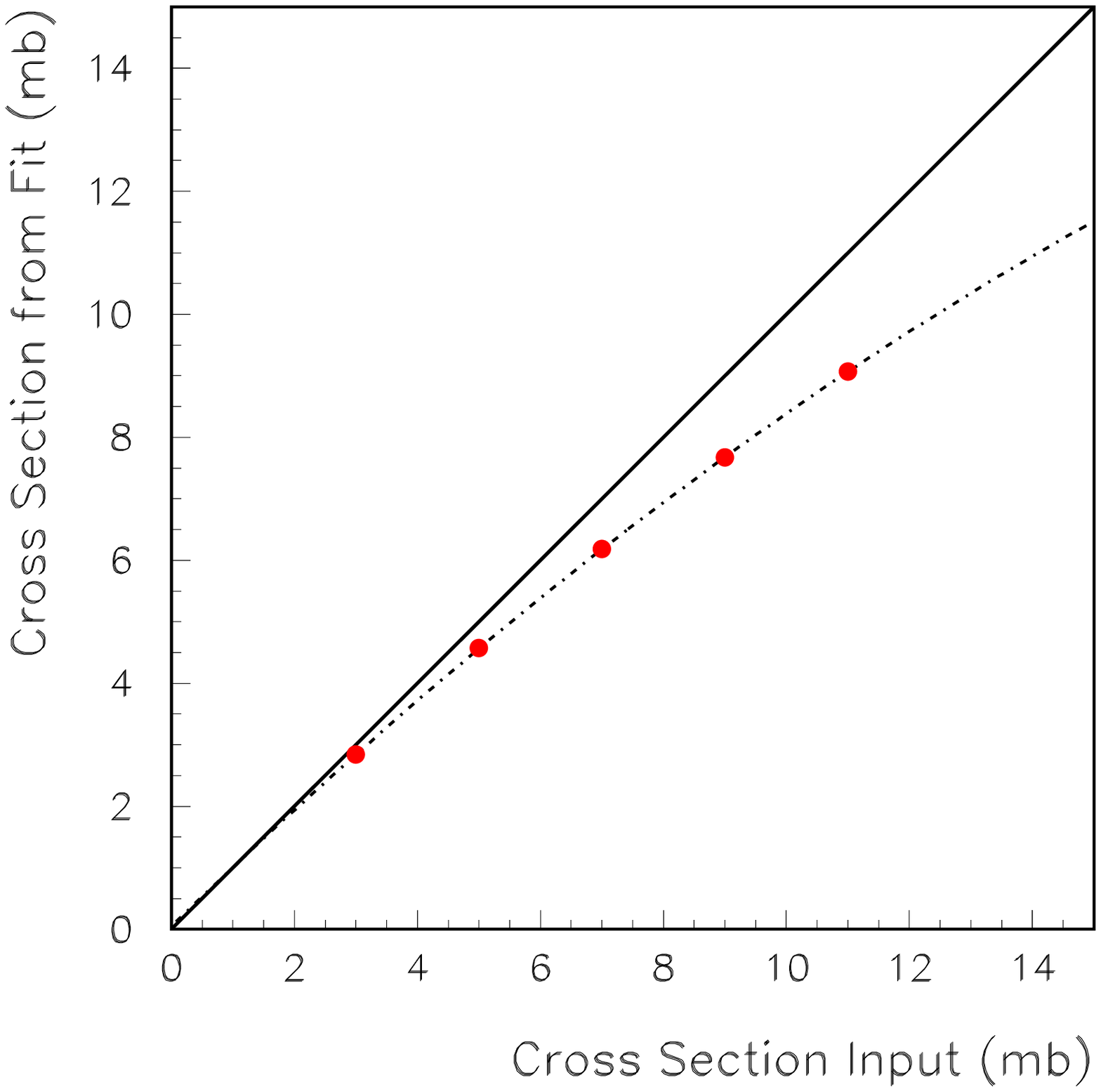,width=4.0in,bbllx=0pt,bblly=140pt,bburx=560pt,bbury=680pt,clip=}
\caption{Breakup cross sections as determined by an exponential fit of $J/\Psi$ yields as a function of $\langle L \rangle$, compared to the actual breakup cross sections used in the calculation.  The value returned from the fit, shown as solid circles for several input values of breakup cross section, are consistently lower that the actual input value (for comparison, a solid line of unit slope is also shown).}
\label{fig:error}
\end{figure}
\end{center}

However, the breakup cross section calculated this way is incorrect, and this error
is straightforward to understand.  For any single 
$J/\Psi$, the survival probability follows an exponential in $L$ for its given 
production point.  However, when calculating the mean survival probability, care 
must be taken on how the mean path length is calculated.  While the calculation 
of $\langle L \rangle$ as described above treats all points in the interaction region 
as contributing equally to production, it does not take into account the fact that, 
due to the exponential nature of absorption, all points in the interaction region 
do not contribute equally to net $J/\Psi$ yields.  In order to accurately determine 
the actual breakup cross section from measured $J/\Psi$ yields, the possible 
production points must be weighted by their survival probability, which depends 
on the breakup cross section; thus, a proper calculation of $\sigma_{J/\Psi-N}$ 
could be accomplished via an iterative calculation, similar in spirit to the one 
detailed here, which exactly returns the input cross section if a weighted mean path 
length is used.  Since surviving $J/\Psi$ particles preferentially come from near 
the outer edges of the interaction region, an exponential fit versus a linearly-averaged 
$\langle L \rangle$ systematically returns an apparent breakup cross section which is 
less than the actual value. In Fig.~\ref{fig:error}, apparent breakup cross sections 
for various sample input breakup cross sections are shown; a fit to the points yields:
\begin{equation}
{\rm \sigma_{actual}=0.118 + 0.932 \, \sigma_{apparent} + 0.030 \, \sigma_{apparent}^{2}}
\end{equation}
The exact values for this fit correspond to the centrality bins chosen, and vary 
slightly for different binning.  With this caveat in mind, for an apparent breakup 
cross section of 6.2$\pm$0.7 mb, as reported in \cite{na50_qm97}, our model 
indicates that the actual breakup cross section is 7.1$\pm$0.9 mb.

\section{Conclusions}
We have calculated the survival probability of $J/\Psi$ particles in various 
colliding systems using a Glauber model, and compared the results to the nuclear 
absorption observed by the NA50 collaboration\cite{na50}.  
While our calculation does not reconcile the issue 
of whether anomalous $J/\Psi$ suppression has been observed in the NA50 Pb+Pb data,
it does indicate that the apparent breakup cross sections determined by a fit of yields 
versus $\langle L \rangle$ are systematically lower 
than the actual value, and a proper evaluation of the breakup cross section 
requires an iterative calculation.  



\end{document}